\documentstyle[prd,aps]{revtex}
\begin{document}

\title{Hyperon Nonleptonic Weak Decays Revisited}

\author{M.D. Scadron} 
\address{Physics Department, University of Arizona,
 Tucson, AZ 85721, USA}
\author{D. Tadi\'{c}}
\address{ Physics Department, Faculty of Science, Zagreb University,
 Bijenicka c.32, 10000 Zagreb, Croatia}

\maketitle

\begin{abstract}
We first review the current algebra - PCAC approach to nonleptonic octet
baryon
14 weak decay B \( \rightarrow  \) \( B^{\prime } \)\( \pi  \) amplitudes.
The needed four parameters are independently determined by \( \Omega
\rightarrow \Xi \pi  \),\( \Lambda K \)
and \( \Xi ^{-}\rightarrow \Sigma ^{-}\gamma  \) weak decays in dispersion
theory tree order. We also summarize the recent chiral perturbation theory
(ChPT)
version of the eight independent B \( \rightarrow  \) \( B^{\prime }\pi  \)
weak \( \Delta I \) = 1/2 amplitudes containing considerably more than eight
low-energy weak constants in one-loop order.
\end{abstract}

\section{Introduction}

For about thirty-five years standard current algebra - partially conserved
axial
current (CA-PCAC) studies have been used to describe hyperon nonleptonic weak
decays. Specifically the measured {[}1{]} seven s-wave B \( \rightarrow  \)
\( B^{\prime } \)\( \pi  \) amplitudes were roughly explained {[}2{]} using
CA-PCAC and SU(3) symmetry for a (d/f)\( _{w} \) ratio of \( \sim  \)-0.4.
Then the observed {[}1{]} p-wave B \( \rightarrow  \) \( B^{\prime }\pi  \)
amplitudes were predicted using a baryon pole model with pseudoscalar
couplings
{[}3{]} for a different \( (d/f)_{w} \) ratio of \( \sim -0.8 \). This s-wave
p-wave mismatch was roughly eliminated when decuplet intermediate state pole
graphs were included {[}4{]}. 

Related techniques were used  to calculate weak strangeness violating vertices,
such as NNK, which were needed in the description of hypernuclear decays.
Thus it might be useful to revisit this particular approach.

More recently, an alternative chiral approach has been used to study B \(
\rightarrow  \)
\( B^{\prime }\pi  \) weak decays, with pseudoscalar - baryon PBB couplings
replaced by pseudovector PBB couplings in a chiral perturbation theory (ChPT)
heavy baryon formulation context {[}5, 6{]}. However the s-wave, p-wave
amplitude
mismatch is still a problem for ChPT {[}5-7{]}. While decuplet \emph{loops}
are folded into the analysis, only the usual low energy constants (LECs) are
considered in leading (non-analytic) log approximation in refs. {[}5, 6{]},
but additional counterterms are included in ref. {[}7{]}. In fact {[}7{]}
accounts
for \emph{all} terms at one loop order, which includes approximately 20
counterterm
terms in their eq. (41), 34 momentum-dependent divergence terms in eq. (44),
18 mesonic lagrangian terms in eq. (B.6), 17 second-order lagrangian terms in
(B.8), 38 second-order lagrangian terms in (B.9), 88 relativistic correction
terms in (B.10), 33 momentum-dependent divergence terms in (B.17), 27 Z-factor
terms in (C.1), 55 terms in the p-wave \( \Sigma ^{+}n \) amplitude in (D.1),
40 terms in the p-wave \( \Sigma ^{-}n \) amplitude (D.2), 35 terms in the
p-wave \( \Lambda p \) amplitude (D.3) and 37 terms in the p-wave amplitude
(D.4).

In this paper we return to the original CA-PCAC picture of B \( \rightarrow
 \)
\( B^{\prime }\pi  \) weak decays as formulated in {[}2, 3{]} and then
resolved
in refs. 4. First in Sec. 2 we review the chiral, \( \Delta I \) = 1/2 and
SU(3) octet structure of B \( \rightarrow  \) \( B^{\prime }\pi  \) weak
amplitudes.
Next in Sec. 3 we fold in the decuplet (D) resonance corrections in a
dispersion
theory context. We also confirm the $<B|H_{w}|D>$ weak scales, h\( _{2}
\),
h\( _{3} \). Finally in Sec. 4 we resolve the long-standing mismatch between
the s-wave and p-wave B \( \rightarrow  \) \( B^{\prime }\pi  \) weak
amplitudes.
We give our conclusions in Sec. 5. In Appendix A we list the various SU(3)
octet
and decuplet pole amplitudes needed for this CA-PCAC analysis. In Appendix B
we list instead the various resonance contributions to nucleon
pion-photoproduction;
the latter demonstrating that the \( \Delta  \) decuplet resonance plays a
dominant role, far more important than \( \frac{1}{2}^{-} \) resonances. The
same pattern should also hold for hyperon nonleptonic weak decays.

\section{Chiral, SU(3) and \protect\protect\protect\( \Delta \protect
\protect \protect \)I
= 1/2 Structure}

The CA-PCAC approach employs the chiral commutator \( [Q+Q_{5}, \) \(
H_{w}]=0 \),
for a weak hamiltonian density built up from V-A currents. Then the resulting
charge commutator amplitude \( M_{cc} \) for \( f_{\pi }\approx 93 \) MeV
satisfies:

\begin{equation}
\label{1}
if_{\pi
}M_{cc}=-<B^{f}|[Q_{5}^{j},H_{w}]|B^{i}>=if^{fjb}<B^{b}|H_{w}|B^{i}>-if^{jia
}<B^{f}|H_{w}|B^{a}>,
\end{equation}  
where the parity-violating (pv) and parity-conserving (pc)
amplitudes (between baryon spinors) are {[}8{]}
\begin{equation}
\label{2}
M=<B'\pi |H^{pv}_{w}+H^{pc}_{w}|B>=iA+\gamma _{5}B.
\end{equation}
 With \( H^{pc}_{w} \) transforming like \( \lambda _{6} \) we invoke {[}9{]}
\( <B'|H^{pv}_{w}|B>=0 \), along with the SU(3) structure {}

\begin{equation}
\label{3}
<B^{f}|H^{pc}_{w}|B^{i}>=h_{1}(d_{w}d^{f6i}+f_{w}if^{f6i}),
\end{equation}
for the scale \( h_{1}\approx 22eV \) \emph{predicted} in
Sec. 4 coupled with \( (d/f)_{w}\approx -0.811 \) and \( d_{w}+f_{w}=1 \),
the former near the fitted octet pc pole amplitude of refs. {[}3,4,10{]}. We
list the various SU(3) scales from (3) in Appendix A. {}

Next to separate off the rapidly varying pole amplitudes \(
M=M^{P}+\overline{M} \)
we follow the soft-pion method (pseudoscalar PBB coupling) and determine the
(slowing varying) background \( \overline{M} \):

\begin{equation}
\label{4}
M=M(0)+M^{P}-M^{P}(0),
\end{equation}
 where \( M(0)=M_{cc} \) from (1) and \( M^{P}(0) \) is
evaluated
at the soft point \( p_{\pi }=0 \). Then the seven pv s-wave amplitudes have
the leading form found from CA-PCAC using (1) and (2):  {}

\begin{equation}
\label{5}
A_{cc}=-\frac{1}{f_{\pi
}}[if^{fjb}<B^{b}|H^{pc}_{w}|B^{i}>-if^{jia}<B^{f}|H^{pc}_{w}|B^{a}>].
\end{equation}
 Also the seven pc p-wave amplitudes are dominated by the
rapidly
varying octet pole terms {[}4{]}: {}

\begin{equation}
\label{6}
B_{8}=-(m_{f}+m_{i})\sum
_{n}(\frac{g^{fn}H_{pc}^{ni}}{(m_{i}-m_{n})(m_{f}+m_{n})}-\frac{H_{pc}^{fn}g
^{ni}}{(m_{n}-m_{f})(m_{i}+m_{n})}).
\end{equation}

 In equation (6) we employ the strong interaction PBB
pseudoscalar
coupling constants \( g^{\pi fi} \) listed in Table I with d/f ratio {[}4{]}
\( (d/f)_{P}\approx 2.1 \) and \( (d+f)_{P}=1 \) so that \( d_{P}=0.678 \),
\( f_{P}=0.322 \). {}

\begin{center}
{\bf Table I: Strong coupling constants}
\vspace{0.3cm}

\begin{tabular}{|c|c|}
\hline 
\( g_{\pi ^{-}pn},g_{\pi ^{+}np} \)&
 \( =\sqrt{2}g\approx 19.0 \)\\
\hline 
\( g_{\pi ^{o}pp}, \) \( -g_{\pi ^{o}nn} \)&
 \( =g\approx 13.4 \)\\
\hline 
\( g_{\pi \Sigma \Lambda } \)&
 =\( \frac{2}{\sqrt{3}}d_{p}g\approx 10.5 \)\\
\hline 
\( g_{\pi ^{o}\Sigma ^{+}\Sigma ^{+}} \), \( g_{\pi ^{-}\Sigma ^{o}\Sigma
^{-}} \),
\( -g_{\pi ^{o}\Sigma ^{o}\Sigma ^{+}} \)&
 \( =2f_{p}g\approx 8.6 \)\\
\hline 
\( g_{\pi ^{o}\Xi ^{o}\Xi ^{o}} \)&
 \( =g(f-d)_{p}\approx -4.8 \)\\
\hline 
\( g_{\pi ^{-}\Xi ^{o}\Xi ^{-}} \)&
 \( \sqrt{2}g(d-f)_{p}\approx 6.7 \) \\
\hline 
\end{tabular}
\end{center}
{}{}\vspace{0.3cm}
However decuplet intermediate states contribute at the 30\% level and we
consider
\( A_{10} \) and \( B_{10} \) amplitudes and couplings in Sec. 3.

Lastly we should fold in \( \Delta I=1/2 \) and SU(3) breaking corrections.
The present s and p wave \( B\rightarrow B'\pi  \) data {[}1{]} listed in
Table
II shows that the \( \Delta I=1/2 \) rules \( \Lambda
_{-}^{o}=-\sqrt{2}\Lambda _{o}^{o} \)
and \( \Xi _{-}^{-}=-\sqrt{2}\Xi _{o}^{o} \) are well satisfied both for s
and p waves. However \( \Sigma _{o}^{+}=(\Sigma _{+}^{+}-\Sigma
_{-}^{-})/\sqrt{2} \)
is violated at the 10\% (20\%) level for s waves (p waves), as is the
Lee-Sugawara
{[}11{]} SU(3) sum rule \( \Sigma _{o}^{+}=(\Lambda _{-}^{o}+2\Xi
_{-}^{-})/\sqrt{3} \).
For simplicity we ignore \( \Delta I=1/2 \) corrections (transforming like
\emph{27}, the group theory language used in refs {[}4{]}) 
but fold in larger SU(3)-breaking \( \Lambda(1405) \)
terms contributing only to \( \Sigma _{+}^{+} \) and \( \Sigma _{-}^{-} \)
amplitudes. Thus the final forms for our s and p wave amplitudes will be

\begin{equation}
\label{7}
A=A_{cc}+A_{10}+A_{\Lambda '},B=B_{8}+B_{10}+B_{\Lambda '}.
\end{equation}

At the quark level, the dominant W-exchange and smaller W self-energy single
quark line diagrams {[}12{]} naturally transform as \( \Delta I=1/2 \), as
first shown by Koerner and Pati and Woo {[}13{]}. Nevertheless our predicted hadron level
amplitudes displayed in Appendix A will have the form of eqs. (7) expressed
in terms of the strong coupling constants of Table I, measured baryon masses
and weak hamiltonian densities as ocurring in eqs. (5) and (6) above.

\newpage

\begin{center}

{ \bf Table II: Observed hyperon decay amplitudes}
\vspace{0.3cm}
\[
\begin{array}{cccc}
 &  & \underline{(s-wave)10^{6}A} & \underline{(p-wave)10^{6}B}\\
(\Lambda _{-}^{o}) & \Lambda \rightarrow p\pi ^{-} & 0.323\pm 0.002 &
2.20\pm 0.05\\
(\Lambda _{o}^{o}) & \Lambda \rightarrow n\pi ^{o} & -0.237\pm 0.003 &
-1.59\pm 0.14\\
(\Sigma _{o}^{+}) & \Sigma ^{+}\rightarrow p\pi ^{o} & -0.326\pm 0.011 &
2.67\pm 0.15\\
(\Sigma _{+}^{+}) & \Sigma ^{+}\rightarrow n\pi ^{+} & 0.014\pm 0.003 &
4.22\pm 0.01\\
(\Sigma _{-}^{-}) & \Sigma ^{-}\rightarrow n\pi ^{-} & 0.427\pm 0.002 &
-0.14\pm 0.02\\
(\Xi ^{-}) & \Xi ^{-}\rightarrow \Lambda \pi ^{-} & -0.450\pm 0.002 &
1.75\pm 0.06\\
(\Xi _{o}^{o}) & \Xi ^{o}\rightarrow \Lambda \pi ^{o} & 0.344\pm 0.006 &
-1.22\pm 0.07
\end{array}\]
\end{center}

\section{Decuplet Weak Corrections}

Recall that the decuplet 33 \( \Delta (1232) \) resonance plays a 30\% role
in low energy \( \pi N \) scattering and it was the origin of the Chew-Low
model {[}14{]} of crossing, analyticity and unitarity, leading to dispersion
relations in particle physics. For the four observed {[}1{]} strong decays
\( \Delta \rightarrow N\pi  \),
\( \Sigma ^{*}\rightarrow \Lambda \pi  \), \( \Sigma \pi  \), and \( \Xi
^{*}\rightarrow \Xi \pi  \),
the strong interaction hamiltonian density is

\begin{equation}
\label{8}
H_{DBP}=\frac{2g_{DBP}}{(m_{D}+m_{B})}p^{B}\cdot
\overline{D}^{(abc)}\epsilon _{cde}B_{b}^{e}P_{a}^{d}\; ,
\end{equation}
with \( g_{DBP}=g_{2}\approx 15.7 \) from data (relative to
\( g_{BBP}=g_{\pi NN}=g\approx 13.4) \). This is consistent with the sign
\( g_{2}g>0 \)
found from photoproduction experiments and from SU(6) currents. Note too that
\( g_{2}\approx 15.7 \) is near Hoehler's narrow \( \Delta  \) width value
\( g_{\Delta N\pi }^{*}\approx 1.83 \) \( m_{\pi }^{-1}\approx 13.1 \) \(
GeV^{-1}, \)
with here \( g_{\Delta N\pi }^{*}=2g_{2}(m_{\Delta }+m_{N})^{-1}\approx14
.5 \)
\( GeV^{-1}. \) This will justify our later neglect of decuplet widths, with
\( g_{2} \) 10\% higher than Hoehler's estimate. {}

For the analogue weak pv, pc transition D\( \rightarrow  \)B\( \pi  \) one
writes {[}4, 15{]}

\begin{equation}
\label{9}
<B\mid H_{w}^{pv,pc}|D>=h_{2,3}\overline{B}_{e}^{b}\epsilon ^{c2e}D_{(3bc)},
\end{equation}
and shortly we will demonstrate in many ways that this
dimensionless
pv weak DB scale is {[}4{]} {}

\begin{equation}
\label{10}
h_{2}\approx -0.2\cdot 10^{-6}.
\end{equation}
However, the similar pc weak DB scale turns out to be a
factor
of 2 greater {[}15{]}: {}

\begin{equation}
\label{11}
h_{3}\sim -0.4\cdot 10^{-6}.
\end{equation}
Then using soft pion or hard pion techniques combined with
dispersion relations (instead using field theory spin 3/2 poles one obtains
additional nonunique parameters A and Z {[}16{]}), the resulting decuplet pole
pv and pc amplitudes \( A_{10} \) and \( B_{10} \) have the form for \(
B_{i}\rightarrow B_{f}\pi  \)
{[}4{]} {}

\begin{equation}
\label{12}
A_{10,}B_{10}=\frac{(m_{B_{i}}\mp m_{B_{f}})}{3}[g_{B_{f}D\pi
}\frac{(m_{D}\pm
m_{B_{i}})}{m_{D}^{2}}H_{DB_{i}}^{pv,pc}+H_{B_{f}D'}^{pv,pc}\frac{(m_{D'}\pm
 m_{B_{f}})}{m_{D'}^{2}}]g_{D'B_{i}\pi }.
\end{equation}

To confirm the \( <B|H_{w}|D> \) weak scales in (9, 10), first one extracts
the \( h_{2} \) scale from \( \Omega \rightarrow \Xi \pi  \) data, by defining
the pc and pv amplitudes as

\begin{equation}
\label{13}
<\pi \Xi |H_{w}|\Omega >=-\overline{u}(\Xi )(E_{pc}+i\gamma
_{5}F_{pv})p^{B}_{\mu }u^{\mu }(\Omega ),
\end{equation}

\begin{equation}
\label{14}
|E_{pc}|\approx (m_{\Omega }+m_{\Xi })^{-1}[24\pi m_{\Omega }^{2}\Gamma
_{\Omega \Xi \pi }/p^{3}]^{1/2},
\end{equation}
giving \( |E_{pc}|\approx 1.33\cdot 10^{-6} \) \( GeV^{-1} \)
from measured \( \Omega ^{-}\rightarrow \Xi ^{o}\pi ^{-} \) weak decay
{[}1{]}.
Then CA-PCAC requires {[}12, 15{]} {}

\begin{equation}
\label{15}
|E_{pc}(\Xi ^{o}\pi ^{-})|\approx |h_{2}/\sqrt{2}f_{\pi }|,|h_{2}|\approx
0.18\cdot 10^{-6}.
\end{equation}
This \( h_{2} \) scale is supported by the \( K^{o} \)
tadpole
pictures of \( |<\pi \pi |H^{pv}_{w}|K^{o}>| \) and \( |<\Xi
^{-}|H_{w}^{pc}|\Omega ^{-}>|=|h_{2}| \),
yielding {[}12, 15, 17{]} {}

\begin{equation}
\label{16}
|h_{2}|\approx |2g_{\Omega \Xi K^{o}}<2\pi ^{o}|H_{w}|K^{o}>f_{\pi
}^{2}/m^{2}_{K}|\approx 0.23\cdot 10^{-6},
\end{equation}
for the coupling \( g_{\Omega \Xi K^{o}}\approx
\sqrt{2}/f_{K}\approx 12.5 \)
\( GeV^{-1} \) and observed amplitude {[}1{]} \( |<2\pi
^{o}|H_{w}|K^{o}>|\approx 26.26\cdot 10^{-8} \)
\( GeV \).  {}

Also the GIM constituent quark model using SU(6) wave functions predicts
{[}17{]}

\begin{equation}
\label{17}
|h_{2}|_{s\rightarrow d}\approx \frac{G_{F}s_{1}c_{1}}{8\pi
^{2}3\sqrt{3}}(\frac{m_{\Omega }+m_{\Xi }}{m_{\Omega }-m_{\Xi
}})(\frac{m_{s}-m_{d}}{m_{s}+m_{d}})\frac{m_{\Omega }}{m_{d}}\approx
0.15\cdot 10^{-6}.
\end{equation}
Given the compatibility between the pv DB scales in (15, 16,
17) we extend the latter quark model picture to estimate the pc DB scales as
{[}15{]} {}

\begin{equation}
\label{18}
|h_{3}|=|\frac{m^{2}_{K}}{m^{2}_{\kappa }}h_{2}\frac{<0|H^{pc}_{w}|\kappa
>}{<0|H^{pv}_{w}|K>}|\sim
|\frac{m_{s}+m_{d}}{m_{s}-m_{d}}h_{2}\frac{m^{2}_{K}}{m^{2}_{\kappa }}|\sim
0.4\cdot 10^{-6}
\end{equation}
for \( m_{s}/m_{d}\approx 1.45 \) and a kappa mass of 850-900
MeV {[}18{]}. {}

With hindsight, always using dispersion theory and unitarity but with
pseudoscalar PBB and PDB couplings, our chiral weak V-A, CA-PCAC scheme will be
renormalized in tree order and NO counterterms will be needed when the
imaginary parts of the weak hyperon amplitudes are evaluated on mass shell via
unitarity (as required using dispersion theory). Contrast this with axial-vector
divergences used for ABB and ADB couplings (needed for the ChPT scheme) which
lead to an unrenormalizable theory involving many counterterms.

\bigskip
\begin{center}
{\bf IV.  DETERMINING THE D/F RATIOS AND 
$<B'|H^{pc}_{w}|B>$ $\textstyle{h}_{1}$ SCALE}
\end{center}

Recall that the strong d/f ratio for axial vector couplings is known to be
\( (d/f)_{A}\approx 1.74 \),
near the SU(6) value 1.50 and the value used for ChPT {[}5-7{]}. However for
the soft pion CA-PCAC method involving pseudoscalar PBB couplings of (6)
needed
for the \( B_{8} \) amplitudes listed in Table I, it was stressed in refs.
{[}4{]} that the strong d/f ratio of \( (d/f)_{P}\approx 2.1 \) is more
appropriate.

Given the observed {[}1{]} fourteen s-wave and p-wave amplitude listed above
in Table II, a good fit in Tables III follow from \( A=A_{cc}+A_{10} \) and
\( B=B_{8}+B_{10} \) (along with the small \( \Lambda '(1405) \) corrections
to the \( \Sigma _{+}^{+} \) and \( \Sigma _{-}^{-} \) amplitudes) developed
in refs. {[}4{]} and in Secs. 2 and 3 and listed in the appendix. This fit
depends
on two additional parameters \( (d/f)_{w} \) and \( h_{1} \) in the pc weak
hamiltonian eq.(3) assuming the values \( (d/f)_{w}\approx -0.811 \) and \(
h_{1}\approx 22 \)
\( eV. \) Thus we must first attempt to deduce these two parameters from data
\emph{other} than from \( B\rightarrow B'\pi  \) weak decays so that the good
fits in Tables III are actually \emph{predictions}.

Specifically we study \( \Omega ^{-}\rightarrow \Lambda K^{-} \) weak decay,
with present decay rate {[}1{]}

\begin{equation}
\label{19}
\Gamma (\Omega ^{-}\rightarrow \Lambda K^{-})\approx \frac{p^{3}}{12\pi
m_{ \Omega}}(E_{\Lambda }+m_{\Lambda })|E(\Lambda K^{-})|^{2}\approx
5.43\cdot 10^{-12}MeV,
\end{equation}
predicting the pc amplitude {[}15{]} {}

\begin{equation}
\label{20}
|E(\Lambda K^{-})|\approx 4.27\cdot 10^{-6}GeV^{-1}.
\end{equation}
Assuming E in (20) to be positive and subtracting off the cc
term \( E_{cc}=\sqrt{3}h_{2}/2f_{K}\approx -1.53\cdot 10^{-6} \) \(
GeV^{-1} \)
together with the small s-channel octet pole term (eq. (17a) in {[}15{]}), the
resulting u-channel octet pole amplitude has the form of the \(
B\rightarrow B'\pi  \)
amplitudes in eq.(3) with \( g_{\Omega \Xi ^{o}\kappa ^{-}}=g_{2}\approx
15.7 \): {}

\begin{equation}
\label{21}
|E(\Lambda
K^{-})|_{u-chan.}=\frac{h_{1}}{2}\frac{3}{2}(f-\frac{1}{3}d)_{w}\frac{2g_{
\Omega\Xi ^{o}K^{-}}}{(m_{\Xi ^{o}}-m_{\Lambda })(m_{\Omega }+m_{\Xi
^{o}})}\approx 5.88\cdot 10^{-6}GeV^{-1},
\end{equation}

\begin{equation}
\label{22}
h_{1}(f-\frac{1}{3}d)_{w}\approx 148.4eV.
\end{equation}
Another constraint on \( h_{1} \) and \( (d/f)_{w} \) follows
from (small) \( \Xi ^{-}\rightarrow \Sigma ^{-}\gamma  \) weak radiative
decay,
since the dominant \( \Delta I=1/2 \) W-exchange quark graph then vanishes
{[}19, 12, 15{]}: {}

\begin{equation}
\label{23}
<\Sigma ^{-}|H_{w}^{pc}|\Xi
^{-}>_{w-ex}=\frac{1}{2}h_{1}(d+f)_{w}=0,(d/f)_{w}\rightarrow -1.
\end{equation}

The still smaller \( \Delta I=1/2 \) single quark line (SQL) s\(
\rightarrow  \)d
graph corresponds {[}20{]} instead to \( (d/f)_{w}=0 \), shifting the net
hadronic
d/f towards {[}4{]} \( (d/f)_{w}\approx -0.88 \) or -0.86 or lower as found
from gluon corrections {[}21{]} to \( (d/f)_{w}\sim -0.8. \) The analog
W-exchange
\( B\rightarrow B'\pi  \) configuration corresponds {[}20{]} to \( \Sigma
_{o}^{+}+\sqrt{3}\Lambda _{-}^{o}, \)
for the s-wave charge commutator amplitudes requiring the combination

\begin{equation}
\label{24}
A_{cc}(\Sigma _{o}^{+})+\sqrt{3}A_{cc}(\Lambda
_{-}^{o})=\frac{h_{1}}{2f_{\pi }}(d+f)_{w},
\end{equation}
the same \( (d+f)_{w} \) structure as in (23). Subtracting
the decuplet s-wave combination \( (0.114\cdot 10^{-6}) \) from the s-wave
data combination from Table II \( (0.233\cdot 10^{-6}) \), the still smaller
cc combination in (24)leads to (for \( f_{\pi }\approx 93 \) MeV): {}

\begin{equation}
\label{25}
h_{1}(d+f)_{w}\approx 22.1eV.
\end{equation}
Combining the constraint equations (22) and (25) we
find {}

\begin{equation}
\label{26}
h_{1}\approx 22.1eV,\; (d/f)_{w}\approx -0.811,
\end{equation}

\begin{equation}
\label{27}
d_{w}=-4.29,\; f_{w}=5.29\, .
\end{equation}
Then we predict from (26, 27) and the \( <B'|H_{w}^{pc}|B>=H_{B'B} \) SU(3)
matrix elements of eq.(3):

\begin{mathletters}
\begin{equation}
\label{28}
H_{n\Lambda
}=-\frac{1}{2}h_{1}\sqrt{\frac{3}{2}}(f+\frac{1}{3}d)_{w}\approx -52.2eV,
\end{equation}

\begin{equation}
H_{p\Sigma ^{+}}=-\sqrt{2}H_{n\Sigma
^{o}}=-\frac{1}{2}h_{1}(f-d)_{w}\approx -105.9eV,
\end{equation}

\begin{equation}
H_{\Lambda \Xi
^{o}}=+\frac{1}{2}h_{1}\sqrt{\frac{3}{2}}(f-\frac{1}{3}d)_{w}\approx 90.9eV,
\end{equation}

\begin{equation}
H_{\Sigma ^{-}\Xi ^{-}}=-\sqrt{2}H_{\Sigma ^{o}\Xi
^{o}}=+\frac{1}{2}h_{1}(f+d)_{w}\approx 11.1eV,
\end{equation}
\end{mathletters}
from which the \( A_{cc} \) and \( B_{8} \) amplitudes listed
in Appendix A and are tabulated in Tables III in the concluding section. {}

Finally, the \( \Sigma _{+}^{+} \) and \( \Sigma _{-}^{-} \) decays tabulated
in Table II also receive small contributions from the \( \Lambda
(1405)=\Lambda ' \)
resonance {[}1{]} with estimated amplitudes enhanced by 25\% relative to
Tables
I, II of ref. {[}4{]} because the present observed rate {[}1{]} of \(
\Lambda '\rightarrow \Sigma \pi  \)
has increased by 25\% in 1998 to 50 MeV. For s and p waves we predict

\begin{equation}
\label{29}
A_{\Lambda '}(\Sigma _{+}^{+})=A_{\Lambda '}(\Sigma
_{-}^{-})=\frac{g_{\Lambda '\Sigma \pi }<n|H_{w}^{pv}|\Lambda '>(m_{\Sigma
}-m_{n})}{(m_{\Lambda '}-m_{n})(m_{\Lambda '}-m_{\Sigma })}\sim 0.08\cdot
10^{-6}
\end{equation}

\begin{equation}
\label{30}
B_{\Lambda '}(\Sigma _{+}^{+})=B_{\Lambda '}(\Sigma
_{-}^{-})=\frac{g_{\Lambda '\Sigma \pi }<n|H_{w}^{pc}|\Lambda '>(m_{\Sigma
}+m_{n})}{(m_{\Lambda '}+m_{n})(m_{\Lambda '}-m_{\Sigma })}\sim 0.14\cdot
10^{-6},
\end{equation}
where \( g_{\Lambda '\Sigma \pi }\approx 0.94 \) and \( H_{n\Lambda
}^{pv,pc}\sim 50eV \)
are the extensions of ref. {[}4{]}.

\setcounter{section}{4}
\section{Conclusion}

In this paper we have summarized the up-dated CA-PCAC chiral approach to
nonleptonic
weak \( B\rightarrow B^{\prime }\pi  \) decays. Section 2 reviews the chiral,
SU(3) and \( \Delta I=1/2 \) structure assuming pseudoscalar PBB couplings
in the lead terms. Section 3 reviews the \( \sim  \)30\% decuplet corrections
for both s and p wave amplitudes. Rather than search for the best fit to the
seven s wave and seven p wave measured \( B\rightarrow B^{\prime }\pi  \)
amplitudes
of Table II, we use additional \( \Omega ^{-}\rightarrow \Lambda K^{-}, \)
\( \Xi \rightarrow \Sigma \gamma  \) and \( \Sigma _{o}^{+}+\sqrt{3}\Lambda
_{-}^{o} \)
data to \emph{predict} (using \emph{no} free parameters) these hyperon decays
in Tables III at the pc scale \( h_{1}\approx 22.1eV \) with \(
(d/f)_{w}\approx -0.811 \).
The Appendix A displays all of the tree-level amplitudes needed.

We contrast this simple 30 year old CA-PCAC approach (using pseudoscalar PBB
couplings) with the more recent and much more complex ChPT one-loop order
scheme
of ref. {[}7{]} (using pseudo-vector couplings). In Table IV we display the
resulting one-loop order \( B\rightarrow B^{\prime }\pi  \) amplitudes and
their associated nonanalytic corrections {[}7{]}. As noted in refs {[}5-7{]},
ChPT appears not to account for both s and p wave amplitudes even though this
approach contains more parameters than the number of amplitudes observed.

In fact in a more recent paper {[}22{]}, the authors of ref. {[}7{]} begin by
stating ``In our recent paper a (ChPT) calculation was performed which
included
all terms at one-loop order. This work suffers from the fact, however, that
at this order too many new unknown LECs (low energy constants) enter the
calculation
so that the theory lacks predictive power.'' Instead in ref. {[}22{]} these
authors study ChPT in tree order, but consider only \( \frac{1}{2}^{-} \)and
the Roper \( \frac{1}{2}^{+} \)resonant states, while now ignoring the
(usually
dominant) \( \frac{3}{2}^{+} \)\( \Delta (1232) \) resonance.

In this paper we also have worked only at (dispersion theory) tree level but
found that SU(3) decuplet states analogous to the \( \Delta (1232) \) play
a major role, as long expected. In Appendix B, we demonstrate that \( \pi N \)
photoproduction data clearly shows that the \( \Delta (1232) \) (and not the
1/2 resonances) is the dominant resonance, suggesting a similar pattern for
nonleptonic weak decays.

With hindsight, other chiral theories based on QCD also lead to a reasonable
picture of hyperon nonleptonic weak decays (except for the mismatch between
s- and p-wave amplitudes {[}23{]}, {[}24{]}). Yet even then the \(
\frac{1}{2}^{+} \)baryon
resonances play a more important role in p-waves than do \( \frac{1}{2}^{-}
\)baryon
resonances play in s-waves {[}25{]} (just as we shall note in Appendix B for
photoproduction chiral amplitudes).

It is worth mentioning that our approach, which succesfully describes hyperon
nonleptonic decays, may be usefully applied in the hypernuclear decay
calculations
{[}26{]}. 

In summary, using on-shell dispersion theory-unitarity techniques, the tree-level
$B \to B^\prime \pi$ weak hyperon amplitudes listed in Apendix A involve no
counterterms (as opposed to ChPT). Moreover such graphs a-priori obey (weak
interaction) chiral symmetry since the current-comutator amplitudes 
of (5) manifestly obey the chiral relations (1) and also the octet (decuplet)
pole terms satisfy the chiral relations (6) and (12), respectively. It is
satisfying that this simple chiral model (involving no free parameters) predicts
14 ($s$- and $p$-wave) amplitudes which reasonably match data [1] as listed in
Tables IIIab.

\vskip 0.5cm


\begin{center}

{\bf Table III a: B\protect\protect\protect\( \rightarrow \protect
\protect \protect \)B\protect\( ^{\prime }\protect
\)\protect\protect\protect\( \pi \protect \protect \protect \)
s wave CA-PCAC predictions \protect\protect\protect\( 10^{6}\protect
\protect \protect \)A}
\vspace{0.3cm}

\[
\begin{array}{cccccc}
 & \underline{cc} & \underline{10} & \underline{\Lambda '} &
\underline{Theory} & \underline{Data}\\
\Lambda _{-}^{o} & 0.397 & -0.092 &  & 0.305 & 0.323\pm 0.002\\
\Lambda _{o}^{o} & -0.281 & 0.065 &  & -0.216 & -0.237\pm 0.003\\
\Sigma _{o}^{+} & -0.569 & 0.273 &  & -0.296 & -0.326\pm 0.011\\
\Sigma _{+}^{+} & 0 & -0.086 & 0.08 & \sim 0 & 0.014\pm 0.003\\
\Sigma _{-}^{-} & 0.805 & -0.486 & 0.08 & \sim 0.40 & 0.427\pm 0.002\\
\Xi _{-}^{-} & -0.691 & 0.223 &  & -0.468 & -0.451\pm 0.002\\
\Xi _{o}^{o} & 0.489 & -0.153 &  & 0.336 & 0.344\pm 0.006
\end{array}\]

{\bf Table III b: B\protect\protect\protect\( \rightarrow \protect
\protect \protect \)B\protect\( ^{\prime }\protect
\)\protect\protect\protect\( \pi \protect \protect \protect \)
p wave CA-PCAC predictions \protect\protect\protect\( 10^{6}\protect
\protect \protect \)
B}

\vspace{0.3cm}

\[
\begin{array}{cccccc}
 & \underline{8} & \underline{10} & \underline{\Lambda '} &
\underline{Theory} & \underline{Data}\\
\Lambda _{-}^{o} & 2.17 & 0.41 &  & 2.58 & 2.20\pm 0.05\\
\Lambda _{o}^{o} & -1.56 & -0.29 &  & -1.85 & -1.59\pm 0.14\\
\Sigma _{o}^{+} & 3.16 & -0.39 &  & 2.77 & 2.67\pm 0.15\\
\Sigma _{+}^{+} & 3.94 & -0.13 & 0.14 & \sim 3.95 & 4.22\pm 0.01\\
\Sigma _{-}^{-} & -0.58 & 0.28 & 0.14 & \sim -0.16 & -0.14\pm 0.02\\
\Xi _{-}^{-} & 1.89 & -0.44 &  & 1.45 & 1.75\pm 0.06\\
\Xi _{o}^{o} & -1.31 & 0.32 &  & -0.99 & -1.22\pm 0.07
\end{array}\]

{\bf Table IV: ChPT one-loop order B\protect\protect\protect\(
\rightarrow \protect \protect \protect \)B\protect\( ^{\prime }\protect
\)\protect\protect\protect\( \pi \protect \protect \protect \)
amplitudes {[}7{]}}

\vspace{0.3cm}

\[
\begin{array}{ccc}
 & \underline{Theory} & \underline{Data}\\
A(\Lambda _{-}^{o}) & 0.333+0.208 & 0.323\pm 0.002\\
A(\Sigma _{+}^{+}) & 0+0 & 0.014\pm 0.003\\
A(\Sigma _{-}^{-}) & 0.437+0.129 & 0.427\pm 0.002\\
A(\Xi _{-}^{-}) & -0.434-0.174 & -0.450\pm 0.002\\
B(\Lambda _{-}^{o}) & -2.59-0.30 & 2.20\pm 0.05\\
B(\Sigma _{+}^{+}) & 0.15+9.61 & 4.22\pm 0.01\\
B(\Sigma _{-}^{-}) & 0.74-0.35 & -0.14\pm 0.02\\
B(\Xi _{-}^{-}) & 0.74+1.02 & 1.75\pm 0.06
\end{array}\]
\end{center}

\section*{Appendix A: SU(3) structure of \protect\protect\protect\(
A_{cc}\protect \protect \protect \),
\protect\protect\protect\( A_{10}\protect \protect \protect \),
\protect\protect\protect\( B_{8}\protect \protect \protect \),
\protect\protect\protect\( B_{10}\protect \protect \protect \).}

We display all s wave and p wave amplitudes A, B contributing to eqs. (7) with
masses \( m_{p} \), \( m_{n} \) denoted by p, n, etc. and \( f_{\pi
}\approx 93 \)
MeV. The predicted weak hamiltonian matrix elements in eqs. (28) and strong
couplings in Table I are always used. Both A and B are weighted by \(
10^{-6} \).

\[
A_{cc}(\Lambda _{-}^{o})=-\sqrt{2}A_{cc}(\Lambda
_{o}^{o})=-\frac{1}{\sqrt{2}f_{\pi }}H_{n\Lambda }\approx 0.397,\]

\[
A_{cc}(\Sigma _{o}^{+})=\frac{1}{2f_{\pi }}H_{p\Sigma ^{+}}\approx -0.569,\]

\[
A_{cc}(\Sigma _{+}^{+})=-\frac{1}{f_{\pi }}(H_{n\Sigma
^{o}}+\frac{1}{\sqrt{2}}H_{p\Sigma ^{+}})=0,\]

\[
A_{cc}(\Sigma _{-}^{-})=\frac{1}{f_{\pi }}H_{n\Sigma ^{o}}\approx 0.805,\]

\[
A_{cc}(\Xi _{-}^{-})=-\sqrt{2}A_{cc}(\Xi _{o}^{o})=-\frac{1}{\sqrt{2}f_{\pi
}}H_{\Lambda \Xi ^{o}}\approx -0.691.\]

\[
A_{10}(\Lambda _{-}^{o})=-\sqrt{2}A_{10}(\Lambda
_{o}^{o})=\frac{h_{2}g_{2}}{3\sqrt{6}}(\Lambda -p)\frac{(\Sigma
^{*+}+p)}{(\Sigma ^{*+})^{2}}\approx -0.092\]

\[
A_{10}(\Sigma _{o}^{+})=\frac{-\sqrt{2}h_{2}g_{2}}{9}(\Sigma
^{+}-p)[\frac{\Delta ^{+}+\Sigma ^{+}}{(\Delta
^{+})^{2}}+\frac{1}{2}\frac{\Sigma ^{*+}+p}{(\Sigma ^{*+})^{2}}]\approx
0.273\]

\[
A_{10}(\Sigma _{+}^{+})=\frac{h_{2}g_{2}}{9}(\Sigma ^{+}-n)[\frac{\Delta
^{+}+\Sigma ^{+}}{(\Delta ^{+})^{2}}-\frac{1}{2}\frac{\Sigma
^{*o}+n}{(\Sigma ^{*o})^{2}}]\approx -0.086\]

\[
A_{10}(\Sigma _{-}^{-})=\frac{h_{2}g_{2}}{3}(\Sigma ^{-}-n)[\frac{\Delta
^{-}+\Sigma ^{-}}{(\Delta ^{-})^{2}}+\frac{1}{6}\frac{\Sigma
^{*o}+n}{(\Sigma ^{*o})^{2}}]\approx -0.486\]

\[
A_{10}(\Xi _{-}^{-})=-\sqrt{2}A_{10}(\Sigma
_{o}^{o})=-\frac{h_{2}g_{2}}{3\sqrt{6}}(\Xi ^{-,o}-\Lambda )[\frac{\Sigma
^{*-,o}+\Xi ^{-,o}}{(\Sigma ^{*-,o})^{2}}+\frac{\Xi ^{*o}+\Lambda }{(\Xi
^{*o})^{2}}]\approx 0.223,-0.216\]

\[
B_{8}(\Lambda _{-}^{o})=-(\Lambda +p)[\frac{g_{\pi ^{-}pn}H_{n\Lambda
}}{(p+n)(\Lambda -n)}-\frac{g_{\pi ^{-}\Sigma ^{+}\Lambda }H_{p\Sigma
^{+}}}{(\Lambda +\Sigma ^{+})(\Sigma ^{+}-p)}]\approx 2.17\]

\[
B_{8}(\Lambda _{o}^{o})=-(\Lambda +n)[\frac{g_{\pi ^{o}nn}H_{n\Lambda
}}{2n(\Lambda -n)}-\frac{g_{\pi ^{o}\Sigma ^{o}\Lambda }H_{n\Sigma
^{o}}}{(\Lambda +\Sigma ^{o})(\Sigma ^{o}-n)}]\approx -1.56\]

\[
B_{8}(\Sigma _{o}^{+})=-(\Sigma ^{+}+p)[\frac{g_{\pi ^{o}pp}H_{p\Sigma
^{+}}}{2p(\Sigma ^{+}-p)}-\frac{g_{\pi ^{o}\Sigma ^{+}\Sigma
^{+}}H_{p\Sigma ^{+}}}{2\Sigma ^{+}(\Sigma ^{+}-p)}]\approx 3.16\]

\[
B_{8}(\Sigma _{+}^{+})=-(\Sigma ^{+}+n)[\frac{g_{\pi ^{+}pn}H_{p\Sigma
^{+}}}{(p+n)(\Sigma ^{+}-p)}-\frac{g_{\pi ^{+}\Sigma ^{+}\Sigma
^{o}}H_{n\Sigma ^{o}}}{(\Sigma _{+}^{+}+\Sigma ^{o})(\Sigma
^{o}-n)}-\frac{g_{\pi ^{+}\Sigma ^{+}\Lambda }H_{n\Lambda }}{(\Lambda
+\Sigma ^{+})(\Lambda -n)}]\approx 3.94\]

\[
B_{8}(\Sigma _{-}^{-})=(\Sigma ^{-}+n)[\frac{g_{\pi ^{-}\Sigma ^{-}\Sigma
^{o}}H_{n\Sigma ^{o}}}{(\Sigma ^{o}+\Sigma ^{-})(\Sigma
^{o}-n)}+\frac{g_{\pi ^{-}\Lambda \Sigma ^{-}}H_{n\Lambda }}{(\Lambda
+\Sigma ^{-})(\Lambda -n)}]\approx -0.58\]

\[
B_{8}(\Xi _{-}^{-})=-(\Xi ^{-}+\Lambda )[\frac{g_{\pi ^{-}\Sigma
^{-}\Lambda }H_{\Sigma ^{-}\Xi ^{-}}}{(\Lambda +\Sigma ^{-})(\Xi
^{-}-\Sigma ^{-})}-\frac{g_{\pi ^{-}\Xi ^{-}\Xi ^{o}}H_{\Lambda \Xi
^{o}}}{(\Xi ^{-}+\Xi ^{o})(\Xi ^{o}-\Lambda )}]\approx 1.89\]

\[
B_{8}(\Xi _{o}^{o})=-(\Xi ^{o}+\Lambda )[\frac{g_{\pi ^{o}\Sigma
^{o}\Lambda }H_{\Sigma ^{o}\Xi ^{o}}}{(\Lambda +\Sigma ^{o})(\Xi
^{o}-\Sigma ^{-})}-\frac{g_{\pi ^{o}\Xi ^{o}\Xi ^{o}}H_{\Lambda \Xi
^{o}}}{2\Xi ^{o}(\Xi ^{o}-\Lambda )}]\approx -1.31\]

\[
B_{10}(\Lambda _{-}^{o})=-\sqrt{2}B_{10}(\Lambda
_{o}^{o})=\frac{-h_{3}g_{2}}{3\sqrt{6}}(\Lambda +p)[\frac{\Sigma
^{*_{+}}-p}{(\Sigma ^{*_{+}})^{2}}]\approx 0.408\]

\[
B_{10}(\Sigma _{o}^{+})=\frac{\sqrt{2}h_{3}g_{2}}{9}(\Sigma
^{+}+p)[\frac{\Delta ^{+}-\Sigma ^{+}}{(\Delta
^{+})^{2}}+\frac{1}{2}\frac{\Sigma ^{*_{+}}-p}{(\Sigma
^{*_{+}})^{2}}]\approx -0.384\]

\[
B_{10}(\Sigma _{+}^{+})=-\frac{h_{3}g_{2}}{9}(\Sigma ^{+}+n)[\frac{\Delta
^{+}-\Sigma ^{+}}{(\Delta ^{+})^{2}}-\frac{1}{2}\frac{\Sigma
^{*_{o}}-n}{(\Sigma ^{*_{o}})^{2}}]\approx -0.132\]

\[
B_{10}(\Sigma _{-}^{-})=-\frac{h_{3}g_{2}}{3}(\Sigma ^{-}+n)[\frac{\Delta
^{-}-\Sigma ^{-}}{(\Delta ^{-})^{2}}+\frac{1}{6}\frac{\Sigma
^{*_{o}}-n}{(\Sigma ^{*_{o}})^{2}}]\approx 0.282\]

\[
B_{10}(\Xi _{-}^{-})=-\sqrt{2}B_{10}(\Xi
_{o}^{o})=\frac{h_{3}g_{2}}{3\sqrt{6}}(\Xi ^{-,o}+\Lambda )[\frac{\Sigma
^{*_{-},o}-\Xi ^{-,o}}{(\Sigma ^{*_{-},o})^{2}}+\frac{\Xi
^{*_{-},o}-\Lambda }{(\Xi ^{*_{-},o})^{2}}]\approx -0.442,-0.447\]

\section*{Appendix B: Pion photoproduction soft pion chiral theorem and the
dominance
of the \protect\protect\( \Delta \protect \protect \) isobar.}

\setcounter{equation}{0} \renewcommand{\theequation}{B.\arabic{equation}}
First
we state the FFR soft-pion theorem {[}27{]}

\begin{equation}
\label{B.1}
\overline{A_{1}}^{(+)}\: (\nu =t=q^{2}=0)=-\frac{g_{\pi NN}\kappa
^{v}}{4\pi m^{2}_{N}}\approx -0.27m^{-2}_{\pi }\: ,
\end{equation}
where \( A_{1}^{(+)} \) is the first isotopic-even pion
photoproduction
background amplitude (weighted by \( \overline{u}_{N^{\prime }} \) \(
\frac{1}{2} \){[}\( \gamma .k, \)
\( \gamma _{\nu } \){]}\( \,  \)\( \gamma _{5}u_{N} \)\( \,  \)). Next we
saturate this background via the \( \frac{3}{2}^{+}\Delta (1232) \), \(
\frac{1}{2}^{+}N(1440) \),
\( \frac{1}{2}^{-}N(1520) \) and \( \frac{1}{2}^{-}N(1535) \) resonant states
{[}28{]}: {}

\[
\overline{A}_{1,\Delta }^{(+)}\, (q\rightarrow 0)=-\frac{g_{\Delta
}^{*}(m_{\Delta }+m_{N})}{6m_{\Delta }m_{N}}(G_{M}^{*}-3G_{E}^{*})\approx
-0.28\, m^{-2}_{\pi }\]

\[
\overline{A}_{1,\, 1440}^{(+)}\, (q\rightarrow 0)=-\frac{g^{\prime
}_{\kappa _{v}^{\prime }}}{2m_{N^{*}}(m_{N^{*}}+m_{N})}\approx 0.03\,
m^{-2}_{\pi }\]

\[
\overline{A}_{1,\, 1520}^{(+)}\, (q\rightarrow
0)=-\frac{g_{N}^{*}(m_{N^{*}}-m_{N})}{16m_{N^{*}}m_{N}}(G^{\prime
}_{E^{v}}-3G^{\prime }_{M^{v}})\approx -0.01\, m^{-2}_{\pi }\]

\begin{equation}
\label{B.2}
\overline{A}_{1,\, 1535}^{(+)}\, (q\rightarrow 0)=-\frac{g_{\kappa
_{v}^{\prime \prime }}^{\prime \prime
}}{2m_{N^{*}}(m_{N^{*}}-m_{N})}\approx -0.01\, m^{-2}_{\pi }
\end{equation}
with \( g_{\Delta }^{*}=g_{\pi N\Delta }^{*}\approx 2.12\, m^{-1}_{\pi } \)
and

\[
G_{M}^{*}=-\frac{1}{e}\left[ \frac{2}{3}\left( \frac{m^{3}_{N}}{m_{\Delta
}^{2}-m^{2}_{N}}\right) \right]
^{\frac{1}{2}}(3A_{\frac{3}{2}}+\sqrt{3}A_{\frac{1}{2}})\approx 3.07\]

\begin{equation}
\label{B.3}
G_{E}^{*}=-\frac{1}{e}\left[ \frac{2}{3}\left( \frac{m^{3}_{N}}{m_{\Delta
}^{2}-m^{2}_{N}}\right) \right]
^{\frac{1}{2}}(A_{\frac{3}{2}}-\sqrt{3}A_{\frac{1}{2}})\approx 0.05,
\end{equation}
where the latter helicity couplings \(
A_{\frac{1}{2},\frac{3}{2}} \)
are listed in the 1980 PDG tables {[}29{]}. The much smaller couplings for the
N(1440), N(1520) and N(1535) are derived or listed in refs. {[}28{]}. {}

Note that the \( \Delta (1232) \) isobar generates the dominant
photoproduction
background resonance term in (B.2), yet the minor N resonant contributions
still
help to sum the four terms in (B.2) to be the precise match of the FFR chiral
result (B.1). This suggests that the analogue SU(3) decuplet resonant terms
together with the SU(3) octet and current algebra (chiral) terms tabulated in
Appendix A above should (and do) match all 14 measured hyperon nonleptonic
weak
decay s- and p-wave amplitudes.


\begin{thebibliography}{10}
\bibitem{8}We use \( i\gamma _{5}=\gamma _{5} \) (Bjorken-Drell) as in text
books by Schweber
or by Scadron.
\end{thebibliography}
\end{document}